\documentstyle[aps,twocolumn,psfig]{revtex}
\input{psfig}

\def\am{angular momentum\ }
\def\momi{moment of inertia\ }

\def\J{{\cal J}}
\def\J2{{\cal J}^{(2)}}
\def\conf{configuration\ }
\def\confs{configurations\ }

\def\qps{quasi particles\ }
\def\qn{quasi neutron\ }
\def\qpr{quasi proton\ }

\begin{document}
\title {Symmetry Breaking by Proton-Neutron   Pairing\\ }
\author{S. Frauendorf$^{(1,2)}$ and J.A. Sheikh$^{(3,4)}$}
\address{
$^{(1)}$Department of Physics, University of Notre Dame, IN 46556, USA\\ 
$^{(2)}$IKH, Research Center Rossendorf, PF 510119, 
D-01314 Dresden, Germany\\
$^{(3)}$Physik-Department, Technische Universitat Munchen,
85747, Garching, Germany\\
$^{(4)}$Tata Institute of Fundamental Research, Bombay,
400 005, India
}

\maketitle

\begin{abstract}
The symmetries of the $t=1$ and $t=0$ pair-fields are different.
The consequences for rotational spectra are discussed. 
For $t=1$, the concept of spontaneous breaking and subsequent
restoration of the isospin symmetry turns out to be important.
It permits us to describe the proton-neutron pair-correlation within
the conventional   frame of pairing between like particles.  
The experimental data are consistent with the presence of a $t=1$
field at low spin in $N\approx Z$ nuclei.
For a substantial $t=0$ field, the 
spectra of even-even and odd-odd $N\approx Z$ nuclei become
similar. The possibility of a rotationally induced
$ J=1$ pair-field at high spin is considered. 
\end{abstract}


\section{Introduction}

The understanding of the proton-neutron pair-correlations is one of
the major goals of the spectroscopy of nuclei with $ N\approx Z$.
The progress in sensitivity achieved with the large
$\gamma$-ray detector arrays combined with mass seperators permits us
to do detailed spectroscopy in the mass 80 and 50 regions. The advent of 
radioactive beams will hopefully allow us to study even heavier $N\approx Z$
nuclei. These experimental opportunities revived the theoretical
activities devoted to the study of the proton-neutron pairing.
\footnote{See A. Goodman's 
contribution to this meeting and \cite{goodman},
which give an overview of the relevant literature.}
What are the consequences
of  proton-neutron pairing for the excitation spectra?
This is an important question, because the excitation spectra and
in particular rotational bands are the  information that come from
$\gamma$ spectroscopy. In this lecture we shall address a more
specific question: What are the consequences of the 
{\bf proton-neutron pair-field} for the rotational spectra?
The restriction to the pair-field has the advantage that its symmetries
show up directly in the spectra, which are qualitatively different for
the various symmetry types.

\section{The pair-field}
 
The pair-field  appears when the mean-field approximation is applied, which
results in the  Hartree-Fock-Bogolubov (HFB)  equations 
\begin{equation}\label{hfb}
{\cal H}'\left( \begin{array}{c} U\\V \end{array}\right)=e'_i \left( 
\begin{array}{c} U\\V \end{array}\right),
\end{equation}

where
\begin{eqnarray}
{\cal H}'=&
\left( \begin{array}{ccc} h'_{ij}+\Gamma_{ij}&
-(\lambda+\lambda_\tau \tau_i)\delta_{ij}& \Delta_{ij}\\
-\Delta_{ij}^{*}&-h'_{ij}-\Gamma_{ij}&
+(\lambda+\lambda_\tau \tau_i)\delta_{ij} 
\end{array}\right),\label{hfbh}&\\
\Gamma_{ij} =& \sum_{kl} \langle ik|v_a|jl\rangle  \rho_{lk},& \label{hfbg}\\
\Delta_{ij} =& {1 \over 2} \sum_{kl} \langle  ij|v_a|kl\rangle  \kappa_{kl},&
 \label{hfbd}\\
\rho   =& V^{*} V^{T},&\label{hfbr}\\
 \kappa     =& V^{*} U^{T}. &\label{hfbt}
\end{eqnarray}
The quantities in the brackets $\langle v_a\rangle $ in (\ref{hfbg}) 
and (\ref{hfbd})
are the antisymmetric uncoupled
matrix elements of the interaction.
In Eq. (\ref{hfbh}), we have introduced the isospin label $\tau =1, -1$ for neutrons
and protons, respectively and rearranged the chemical potentials 
$\lambda_n$ and $\lambda_p$ which constrain $N$ and $Z$ into 
$\lambda=(\lambda_n+ \lambda_p)/2$ and 
$\lambda_\tau=\lambda_n- \lambda_p$ which fix mass $A$ and the 
isospin projection $T_z$, respectively.  
The HFB solutions are obtained by solving the equations (\ref{hfb})
-(\ref{hfbt})
self-consistently.

The pairs of states  $\{ ij\}$ that define the pair-field (\ref{hfbd})
can be rewritten in a coupled representation  as $\{ t,t_z,i,j\}$, 
which explicitly indicates the isospin $t$ and $t_z$ and  $i ,j$
denote all quantum numbers except the isospin.
If $t=0$, the pair-field is an isoscalar and for $(t=1,t_z)$ it is
an isovector.
The proton-proton (pp) pair-field has $(t=1, t_z=-1)$ and the neutron-neutron
(nn) has $(t=1, t_z=1)$. There are two proton-neutron (pn)
 pair-fields with
$(t=1, t_z=0)$ and $(t=0, t_z=0)$. We use the lower case letters 
$t$ and $t_z$ for the isospin
of the pair-field in order to 
avoid confusion  with the isospin of the states, which we denote
by $T$ and $T_z$. 
Let us restrict to the simple case of one 
$j$-shell, for which the symmetries become most obvious.
 The generalization to many $j$-shells is straight forward. 
 Then the field is specified by $i,j=J,M$, where
 $J,M$ label the angular momentum of the pair. 
Anti-symmetry implies that for $t=1$ the angular momentum $J$ is
even and for $t=0$ it is odd.  The monopole 
$t=1,J=0$ and the dipole  $t=0,J=1$ are the most
important pair-fields because they have the largest
matrix elements for a  short range interaction as well as a
sufficient  number of states available to built up correlations.
\footnote{The $t=0,~J=2j$ field has also a large matrix element
but there is only one such state in the $j$-shell. It may built up
correlations in the space of many $j$-shells. Since its symmetries 
are the same as for the $J=1$ field, we shall not consider it further.}

In the following we assume that the pair-field is either $t=1,~J=0$
or  $t=0,~J=1$. This is the most common solution of the HFB equations.
But coexistence of $t=1$ and $t=0$ fields is possible, as demonstrated by
A. Goodman \cite{goodman}.

\section{The $t=1$ pair-field}

We shall proceed with the assumption that for $40<A<80$ 
the $N\approx Z$ nuclei 
have a $t=1$ pair-field at low spin. This assumption is supported by the 
experiments with two-particle transfer reactions on these nuclei. 
As reviewed  by Bes, Broglia, Hansen and Nathan \cite{pairvib},
the observed cross section can be very convincingly be interpreted 
in terms of a $t=1$ pair-field.  A dynamic picture
of the field is used and multi-boson excitations are considered. 
We shall stay within the static HFB approach and consider the consequences of
the $t=1$ pair-field for the energy spectra. The following discussion 
is based on the material in ref.\cite{pncsm}.

The pair-potential $\Delta$ has the same symmetries as the pairing-tensor
$\kappa$, because the interaction is an invariant with respect to the symmetry
operations. For $t=1$ monopole pairing we have
\begin{eqnarray}\label{t1field}
\vec\kappa= \sum_m \langle j,m,j,-m|00\rangle&& \nonumber\\
\times\left(\begin{array}{c}\langle c^+_{jm,n}c^+_{j-m,n}\rangle\\
\frac{1}
{\sqrt{2}}\langle c^+_{jm,n}c^+_{j-m,p}+c^+_{jm,p}c^+_{j-m,n}\rangle\\   
\langle c^+_{jm,p}c^+_{j-m,p}\rangle
\end{array}\right)&
\begin{array}{c}t_z=1\\t_z=0\\t_z=-1\end{array}.&
\end{eqnarray}
It is a vector in isospace in spherical representation. 
The pair-potential $\vec \Delta$ can be written in an analogous way.    
The original two-body Hamiltonian conserves the isospin (we neglect the 
Coulomb interaction). It is a scalar with respect to rotations in isospace.
We consider the case $N=Z$. Then $\lambda_\tau=0$.
The mean field Routhian ${\cal H}'$ (cf. (\ref{hfbh})) consists of
isoscalar terms, except the pair-potential, which is an 
isovector. Therefore, it breaks the isospin symmetry spontaneously 
\footnote{A discussion of the concept of spontaneous symmetry breaking 
is given in \cite{ringschuck}.}

\subsection{Spontaneous breaking of the isospin symmetry by 
the $t=1$ pair-field}\label{sbiso.sec}
 
Before discussing the  breaking of the symmetry by the isovector pair-field, it is
useful to state the familiar case of spontaneous breaking of the 
spatial isotropy by a  mean-field solution with a deformed density
distribution (c.f. ref.\cite{ringschuck,bm2}). Since the two-body
Hamiltonian is isotropic, this symmetry is broken spontaneously. 
There is a family of mean-field solutions with the same energy which 
correspond to different orientations of the density distribution.
All represent one and the same intrinsic quasiparticle configuration,
which is not
an eigenfunction of the total angular momentum. Any of these solutions can be 
chosen as the intrinsic state. The principal axes of its density
distribution define the body-fixed coordinate system.
The states of good angular-momentum are superpositions of these states of
different orientation, the weight being given by the Wigner
$D$-functions. Thus, the relative importance of the different orientations
is fixed by restoring the angular-momentum symmetry. At the simplest level
of the cranking model, which is valid for sufficiently strong symmetry 
breaking, 
the energy of the good angular-momentum
state is given by the mean-field value.      

\begin{figure}[t]
\mbox{\psfig{file=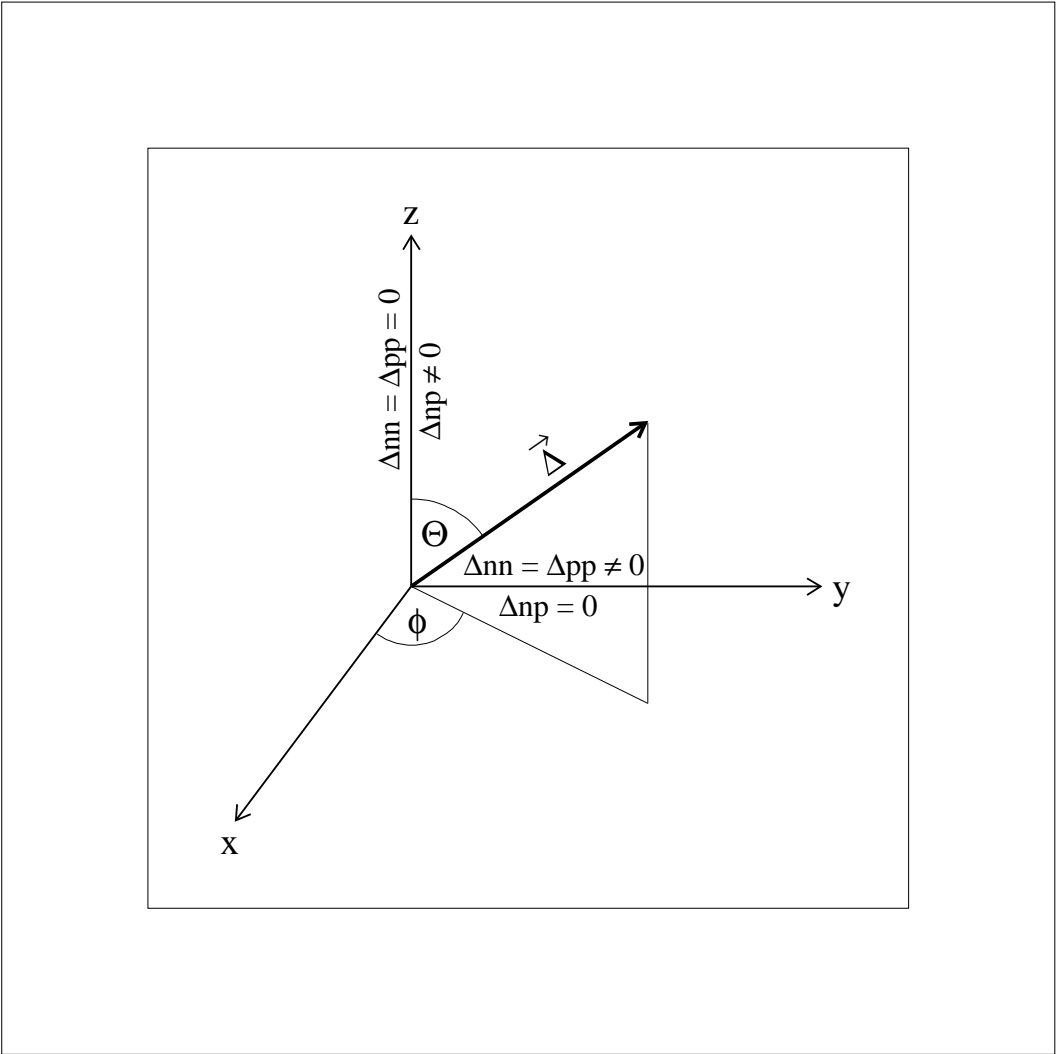,width=14cm}}
\caption{\label{delta.fig}
The isovector pair-field $\vec \Delta$.}
\end{figure}

Let us now consider a $t=1$ HFB solution found for the $N=Z$ system.
The $t=1$ pair-field $\vec \Delta$ is a vector  that
points in a certain direction in isospace,  breaking the 
isospin symmetry. Since the two-body Hamiltonian is isospin invariant, 
the symmetry is a spontaneously broken and all orientations of the 
isovector pair-field : 
\begin{eqnarray}
\Delta_{J,M,t=1,t_z=\pm 1}=\mp \Delta_{J,M,t=1}
 \sin{\theta}\exp{\mp i \phi}/\sqrt{2} \nonumber \\
\Delta_{J,M,t=1,t_z=0}=\Delta_{J,M,t=1} \cos{\theta}
\end{eqnarray}
are equivalent. Fig. \ref{delta.fig} illustrates this family of  HFB 
solutions, the energy of which 
does not depend on the orientation of the pair-field.
In particular, the cases of a pure pn- field ( z-axis) and 
pure pp- and nn-pair-fields ( y-axis) represent
the same intrinsic state. Hence, at the mean-field level
the ratio between the strengths of  pp-, nn- and pn- pair-fields is given by
the orientation of the pair-field, which is not determined by the HFB
procedure.
The relative strengths 
of the three types of pair-correlations becomes
only definite when the isospin symmetry is restored.
The symmetry breaking by the isovector pair-field has been discussed 
before in  \cite{camiz,ginocchio}, where references to
earlier work can be found. 

The most simple way to restore the symmetry is the above
discussed rotor limit, which
assumes that the  rotation is slow as compared with the intrinsic
motion. For the isospace it was discussed in \cite{pairot}.
In essence it amounts to pick one of the orientations of $\vec\Delta$
and call this the intrinsic state. The direction of the y-axis,
corresponding to $\Delta_{np}=0$, is particular useful, 
as will be discussed below.
The symmetry conserving wavefunction is
a product of this intrinsic state and   Wigner $D$-function, which is
the probability amplitude of the different orientations of the intrinsic
state in isospin. For $T=0$ states it is a constant. All orientations of
$\vec \Delta$ are equally probable, 
corresponding to an equal amount of pn-, pp- and nn-
correlation energy. 
In this way the pn- pair-field reappears
via restoration of the isospin symmetry, although the intrinsic state
has only the pp- andd nn- pair-fields.  For states $T_z=T$, the $D$-function
becomes more and more peaked in the y-direction, i. e.  with increasing $T$
the orientations with $\Delta_{pn}\not=0$ become less and less probable.

This simple pair-rotational scheme  is
completely analogous to the familiar rotational bands.  
The intrinsic excitations represent the $T=0$ states. The energies of states
with larger isospin are 
\begin{equation}\label{epairot}
E(T)=E(T=0)+\frac{T(T+1)}{2{\cal J}_T}.
\end{equation}     
The moment of inertia for the isorotation can be calculated by means of 
the cranking procedure. One solves the HFB eqs. for a finite "frequency"
$\lambda_\tau$ and calculates 
\begin{equation}\label{jpairot}
{\cal J}_T=\frac{\langle T_z\rangle}{\lambda_\tau}.
\end{equation}     
The moment of inertia is approximately proportional to the level density at
the Fermi surface. Realistic interactions or shell model potentials are
tuned such that the experimental level density is well reproduced. 
Hence, the pair-rotational energy, which is a combination of symmetry and
the Wigner terms of the binding energy, is expected to be reproduced well.   

\subsection{Intrinsic excitations}\label{t=0.sec}

Like in the case of spatial rotation, the intrinsic excitations are constructed
from the quasi particles (qps) belonging to {\em one} of the 
orientations of pair-field.
We choose the y-direction, $\Delta_{nn}=\Delta_{pp}, \Delta_{np}=0$. 
This is a particularly  convenient choice because it permits to
reduce the construction
of the qp excitation spectrum to the familiar case with no pn-pairing 
\cite{bf79}. The choice of the qp operators is not unique \cite{ginocchio}. 
We choose them to be pure quasi neutrons or quasi protons and 
denote their creation operators   by  $\beta_{t_z,k}^+$,
where $t_z$  indicates  the isospin projection. They are pairwise degenerate,
i.e. the qp Routhians $e'(\omega)_{\frac{1}{2},k}=e'(\omega)_{-\frac{1}{2},k}$
are equal.  Our choice of the orientation of the intrinsic state
has the advantage that its symmetries become obvious. 
Since
\begin{equation}
[ e^{-i\pi Z},{\cal H}'] = [ e^{-i\pi N},{\cal H}'] =0,
\end{equation}
proton and neutron  number parities  are conserved. 
That is states with even or  odd $N$ are different 
\qn \confs and states with even or  odd $Z$ are different 
\qpr \confs. 
The HFB vacuum state has $N$ and $Z$ even: 
Configurations with an odd or even number of quasi neutrons belong to the 
odd or even  $N$, respectively, and the same holds for the protons.
In particular, the lowest $T=0$ state in odd-odd $N=Z$ nuclei is a 
two-qp excitation and different from the ground state of its even-even
neighbor, which is the vacuum. 

However, not all qp configurations are permitted. If $\lambda_\tau=0$,
the qp Routhian   commutes with $T_y$ 
\begin{equation}
[ T_y,{\cal H}'] = 0
\end{equation}
This implies that the qp configurations
have  $T_y$ as a good quantum number. Since $T \geq T_y$, only
 configurations with $T_y=0$ are permitted. 
The detailed discussion of this restriction can be found in \cite{pncsm}.

\subsection{Comparison  with the exact shell model solutions}\label{csmsm.sec}
As a study case, we used
the deformed shell model  Hamiltonian which 
consists of a cranked deformed
one-body term, $h^\prime$ and a scalar two-body delta-interaction 
\cite{srn90,she90,fsr94}. The one-body term is the familiar
cranked-Nilsson mean-field potential which takes into account of the
long-range part of the nucleon-nucleon interaction. The residual short-range
interaction is specified by the delta-interaction,
\begin{equation} \label{H}
H^\prime = h^\prime -g \delta(\hat r_1 - \hat r_2)
\end{equation}
where,
\begin{equation}\label{h'}
h^\prime = -4 \kappa { \sqrt { 4 \pi \over 5 }} Y_{20}  - \omega J_x
\end{equation}
We use 
$G=g\int R_{nl}r^2 dr$ as our energy unit and the deformation
energy $\kappa$ and is related to the deformation parameter $\beta$.
We have diagonalized the Hamiltonian (\ref{H}) exactly for neutrons and protons
in the $f_{7/2}$ shell, for which $\kappa$=1.75 approximately
corresponds to $\beta=0.16$. 
In addition  to its invariance
with respect to rotations in isospace
 (1) is invariant with respect to ${\cal R}_x(\pi)$,
 a spatial rotation
about the x-axis by an angle of $\pi$. As a consequence, the signature $\alpha$
is a good quantum number \cite{ringschuck,bf79}, which implies that the shell model solutions
represent states with the angular momentum $I=\alpha +2n$, $n$ integer.

\begin{figure}[t]
\vspace*{-3cm}
\mbox{\psfig{file=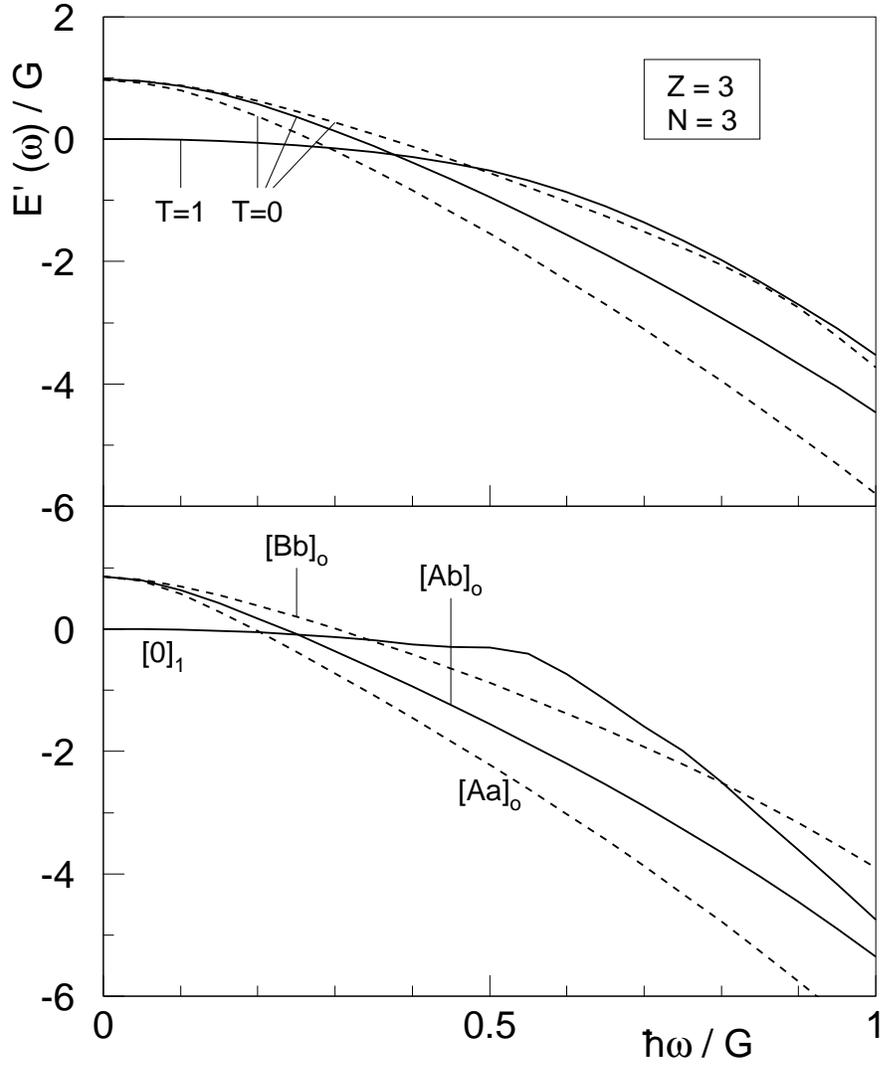,width=14cm}}
\caption{\label{e33.fig}
Total routhians for the $(Z=N=3)$ system. The upper panel shows
the shell model results and the lower the CSM approximation.
Full lines correspond to even spins and dashed ones to odd spins. 
The labeling of the quasiparticle configurations is 
explained in the text. }
\end{figure}

The exact energies obtained by diagonalizing the shell model
Hamiltonian (\ref{H}) for $(Z+N=3+3)$ particles in 
the $f_{7/2}$
shell are shown in the upper panel of figs.
\ref{e33.fig}. The states are classified with respect to 
the isospin and the signature.

We have solved the HFB equations
self-consistently for (4+4) at $\omega=0$. 
The solution is  a $t=1$ pair-field. 
In order to construct mean field solutions for finite 
frequency we adopted the Cranked Shell Model (CSM) approach \cite{bf79}. 
The fields $\Gamma$ and $\Delta_{nn}=\Delta_{pp}$
determined for $\omega=0$ are kept fixed for all other values of $\omega$.
They are also used to describe the (3+3) and (3+4) systems, for which only
$\lambda$ and $\lambda_\tau$ are adjusted 
to have $\langle  N \rangle  = N$ and $\langle  Z \rangle  = Z$ at $\omega=0$.
Earlier studies of a small number of particles in a $j$-shell
showed that the CSM gives better agreement with the exact shell model than
than demanding full self-consistency for all $\omega$  \cite{srn90}.

\begin{figure}[t]
\mbox{\psfig{file=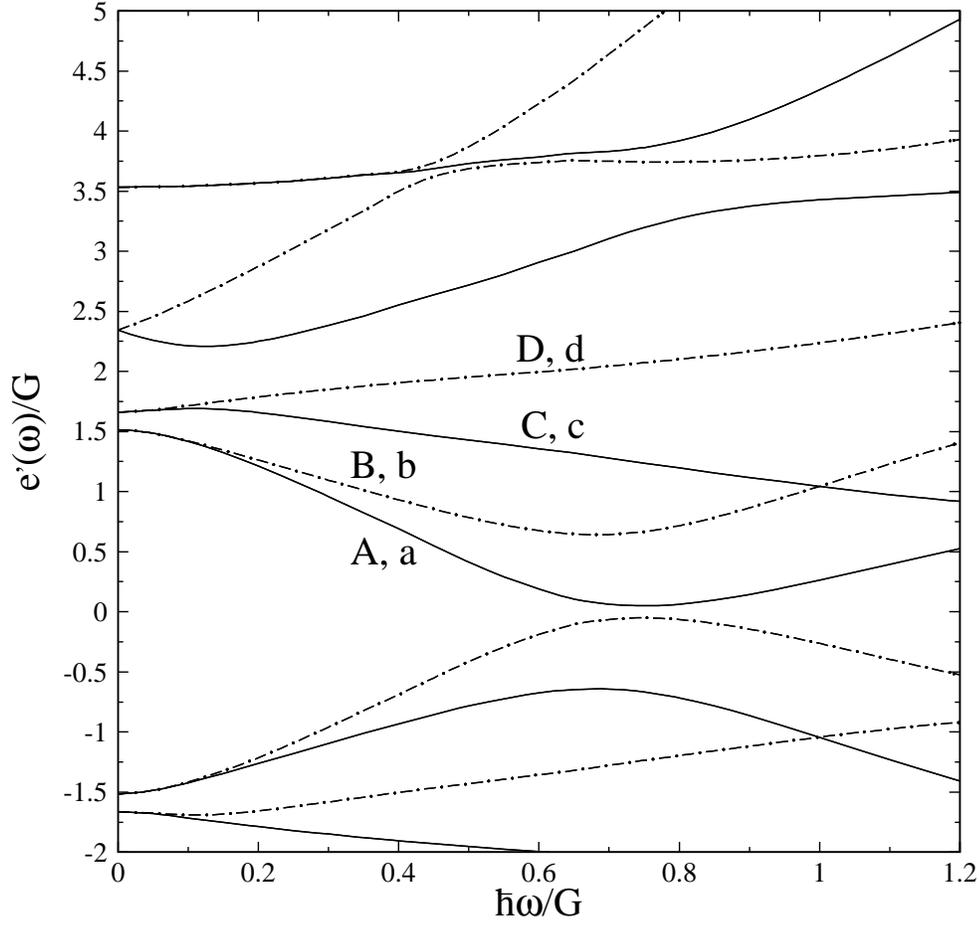,width=14cm}}
\caption{\label{qp.fig}
Quasiparticles in the $f_{7/2}$ shell as function of the rotational
frequency $\omega$. The chemical potential corresponds to a half filled
shell $\langle Z\rangle =\langle N\rangle =4$. The mean-field is kept fixed to the values
calculated  by solving the HFB equations (\protect \ref{hfb}) for 
$\omega=0$.  Full drawn and 
dashed dotted lines denote the favored and unfavored signature ($\alpha =-1/2$
and 1/2 for $f_{7/2}$), respectively.}
\end{figure}

\begin{figure}[t]
\mbox{\psfig{file=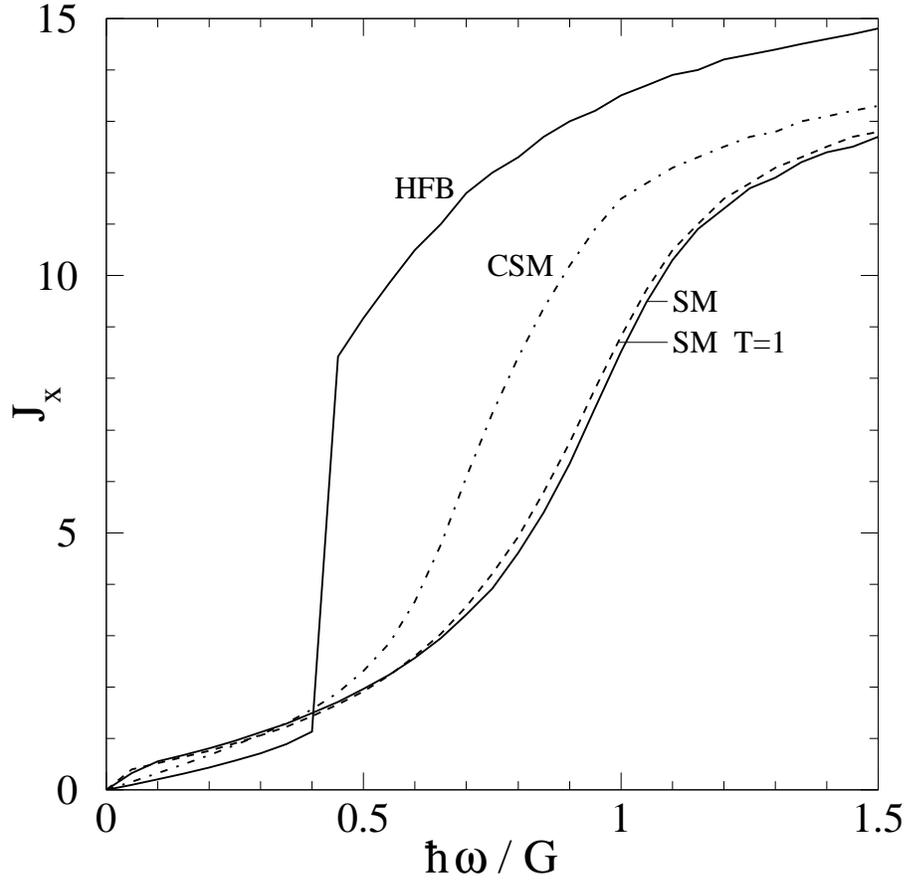,width=14cm}}
\caption{\label{jx.fig}
Angular momentum expectation value $\langle J_x\rangle $ for the yrast-band in 
the $(Z=N=4)$ system. The full shell model result is denoted by SM, 
the  shell model result with a modified two-body interaction leaving out 
the $T=0$ components 
of  the $\delta$-force by SM T=1, the fully selfconsistent HFB calculation
 by HFB and
the CSM approximation by CSM.}

\end{figure}

Fig. \ref{qp.fig} shows the quasiparticle Routhians $e'_i(\omega)$.
All are  two-fold degenerated. They correspond to a quasi proton
and a quasi neutron , which are labeled, respectively, by a, b, c, ...
and A, B, C, ..., adopting  the popular CSM letter convention.
The configurations are constructed by the standard qp occupation scheme, as
described in ref. \cite{bf79}. 
The vacuum [0] corresponds to all negative qp orbitals filled.  
It has  has signature
$\alpha =0$, even $N$ and $Z$ and $T_y=0$. It represents the even-spin $T=0$ 
yrast band of the $(N=Z=4)$  system. The AB-crossing at $\omega=0.6$
corresponds to the {\em simultaneous} alignment of a proton- and a neutron-
pair (because the Routhians are degenerate).
Fig. \ref{jx.fig}  demonstrates that the CSM approximation 
describes the double alignment fairly well.
It also shows  a shell model
calculation where we took off all the $t=0$ components of the delta interaction.
The crossing appears at almost the same frequency as  in the calculation
with the full interaction. Hence the possible $t=0$
 correlations cannot influence
the crossing in an important way, as has been speculated \cite{kr72}.

The lowest two-qp excitation is generated by putting one quasi proton and
one quasi neutron on the lowest  Routhian. We denote
this configuration by $[A,a]_0$.
It has $T_y=0$ and thus correspond to a $T=0$ band. 
The subscript indicates the isospin $T$ of the configuration.
The total signature is $\alpha=1$ and corresponds to an odd-spin band.
The particle numbers
$N$ and $Z$ must be odd, because exciting one quasi neutron 
 changes $N$ from even to 
odd or from odd to even and the same holds for the quasi protons. 
Thus  $[A,a]$
 is the lowest $T=0$ odd-spin band in the 
odd-odd $N=Z$ system. Fig. \ref{e33.fig}
shows the CSM estimate for this band.
The configuration $[B,b]$ is the second odd-spin $T=0$ band and
$[A,b]$ the first even-spin $T=0$ band in the odd-odd system.      
 The configuration $[a,B]$ 
does not generate a new state, because
  only the superposition $([A,b]-[a,B])/\sqrt{2}$ corresponds $T_y=0$,
the other must be discarded. 
To keep the notation simple, we label the configuration as $[A,b]$. But it is
understood that the superposition is meant.
The lower panel of fig. \ref{e33.fig} shows these
\confs, which represent the  three lowest  $T=0$ bands. 

The lowest $T=1$ band 
is  pair-rotational level based  intrinsic vacuum  state, which we denote by
$[0]_1$. Its energy is given by (\ref{epairot}) and the \momi 
${\cal J}_T$   
is found by ``cranking in isospace'' according to (\ref{jpairot}). 

The comparison with the shell model calculation  in fig. \ref{e33.fig} 
demonstrates that the simple pair-rotational scheme 
reproduces  well the position of the $T=1$ even-spin band 
relative to the three lowest $T=0$ bands, the relative position of
which is also well
reproduced by the CSM.  The appearance of the 
$T=1$ even-spin band below  the $T=0$ bands
is a specific feature of the $Z=N$ system.( In   odd-odd nuclei
with $N\gg Z$
all  bands start with an energy larger than $2\Delta$.) Its low energy 
for  $\omega=0$ has the consequence that 
the  $T=1$ even-spin band is crossed by
the aligned odd-spin $T=0$ band. This crossing has been observed in 
$^{74}$Rb \cite{rb74}. 
The similar energy of the $T=1$ and $T=0$ states at $\omega=0$ appears
 as a cancellation between the pair-gap and the isorotational
energy. The configuration
$[Aa]_0$ is shifted by $2\Delta$ with respect to
 the qp vacuum $[0]_0$. The configuration $[0]_1$ is shifted
by $T(T+1)/2{\cal J}_T$. Both quantities are 
nearly equal. This is not a special feature of the $j$-shell model,
as discussed now.

\subsection{Realistic nuclei}

The  energy difference between the lowest $T=0$ and $T=1$ states in 
odd-odd $N=Z$ nuclei has recently by studied by P. Vogel \cite{vogel}
and A. Macchiavelli {\em et al.}\cite{macchiavelli}. It turns out to be
few 100$keV$ for all nuclei with $A>22$, except $A=42$ and 46.
The small difference is an indication for the presence of the $t=1$
pair-field: The $T=0$ state lies at $2 \Delta$ because it is a two-qp
excitation. The $T=1$ state lies at $1/{\cal J}_T$ because it is the 
zero-qp state but the first excited state of the pair-rotational band.
Since the two terms are about the same the two states have almost equal
energy. One may derive experimental values for  $\Delta$ from the odd-even
mass differences  and for    $1/{\cal J}_T$ from the experimental
energies $E(T_z,A)$ within a isobaric chain. These independently determined
values are consistent with the experimental energy differences
$E(T=1)-E(T=0)$ observed in the odd-odd nuclei \cite{vogel,macchiavelli}. 

For the even-even $N=Z$ nuclei the $T=0$ state is the vacuum.
Then lowest $T=1$ state must be a two-qp excitation, which lies
at $2\Delta+ 1/{\cal J}_T$, which is about 5$MeV$. The experimental energy
of the lowest  $T=1$ state agrees well with this estimate \cite{vogel}.

The fact that $2\Delta\approx 1/{\cal J}_T$ 
holds  not only the experimental values but also for the single $j$-shell
model points to a general feature, which remains to be understood.

Since we the orientation of $\vec \Delta$ with $\Delta_{pn}=0$
is a legitimate choice for  the intrinsic state, 
the analysis of rotational bands in realistic nuclei 
can be carried out along the familair scheme without a pn- pair-field. 
One has only to take into account the possibility of low lying 
pair-rotational states and the exclusion of states due to the 
condition $T_y=0$. 
This sheds light on the results of the recent analyzes of high spin
data in nuclei with  $T_z=1/2$ and 1  by means of 
this conventional approach  \cite{kr74,rb75}, which 
find good agreement between theory and experiment.
For  $T_z=1/2$  the first excited pair-rotational state has $T=\frac{3}{2}$. 
It lies at least $\frac{3}{2}(\frac{3}{2}+1)/1(1+1)\approx 1.8$ times higher
than in the nuclei with  $T_z=0$, where it has $T=1$. 
In the $T_z=1$ nuclei it lies even
higher. Thus the lowest bands are  only the intrinsic excitations,
which can be described as qp excitations with $\Delta_{pn}=0$. 
These results support our suggestion that in the
investigated nuclei with $70<A<80$ there is strong $t=1$ pairing.
In this connection we want to point out once more that
 the fact that the mean field theory without an explicit 
pn-  pair-field works well does by no means imply that there is no
 $t=1$  pn- field. On the contrary, it must have a strength comparable
with the pp- and nn- fields in order to restore the isospin symmetry.

\begin{figure}[t]
\vspace*{-3cm}
\mbox{\psfig{file=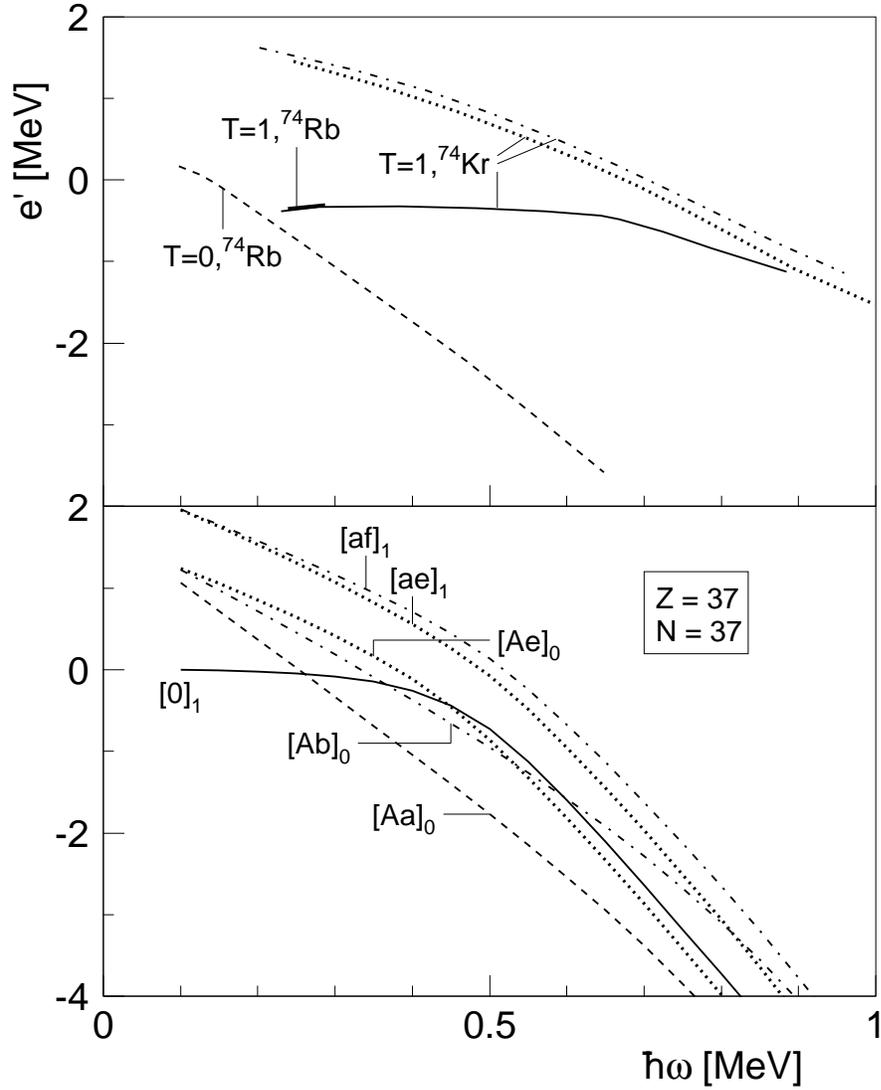,width=14cm}}
\caption{\label{rb74.fig}
Total routhians for  $^{74}_{37}$Rb$_{37}$. The upper panel shows
the experimental routhians \protect\cite{rb74}
 and the lower the CSM approximation. For $T=1$ also the isobaric
analog $T_z=1$ bands in $^{74}_{36}$Kr$_{38}$  \protect\cite{kr74}
are shown. 
The parity and signature assignments $(\pi,\alpha)$ are:
Full lines (+,0), dashed (+,1), dashed dotted (-,0) and
dotted (-,1). A Harris reference is subtracted.
   }

\end{figure}

Fig. \ref{rb74.fig} displays the spectrum of  $^{74}_{37}$Rb$_{37}$.
The upper panel also shows   the $T_z=1$ bands measured in
$^{74}_{36}$Kr$_{38}$. They are  isobaric analog to the $T=1$ bands in 
$^{74}$Rb and should give a good estimates of these  bands.
 Since the $T=1$ states belong to an isobaric triplets, 
we set the ground state energy  of $^{74}$Kr equal to the energy of the 
$I=0$ state in $^{74}$Rb.

\begin{figure}[t]
\mbox{\psfig{file=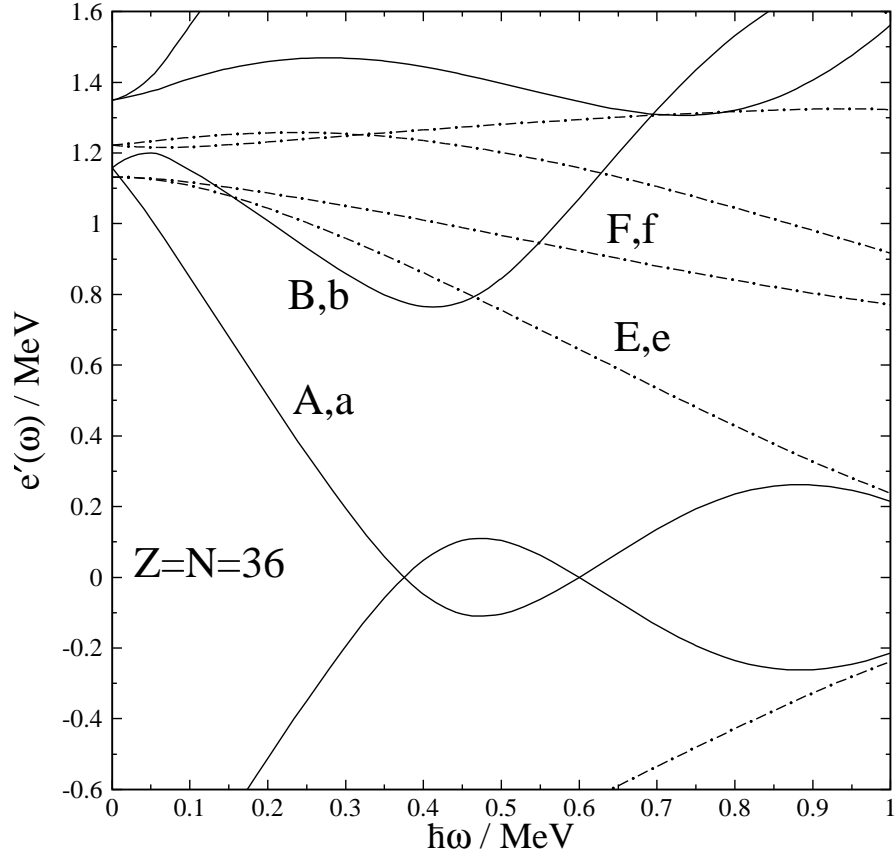,width=14cm,angle=-90}}
\caption{\label{qp72.fig}
Quasiparticles for  $(N=Z=36)$  as function of the rotational
frequency $\omega$.  The mean-field is
the modified oscillator with the deformations $\varepsilon=0.3$,
$\varepsilon_4=0$ and $\gamma=0$ and $\Delta_n=\Delta_p=1.1 MeV$.
The diagram is relevant for both protons and neutrons. 
Full drawn and 
dashed dotted lines denote positive and negative parity, respectively.
The signature is indicated by the letters: $\alpha=1/2$ for A,E and
$\alpha=-1/2$ for B,F. }
\end{figure}

Fig. \ref{qp72.fig} shows the quasi neutron  Routhians for $N=36$, which are
nearly identical with the quasi proton ones.
The standard letter coding is used to label the qp Routhians. The use of
A, a, ...  indicates that the diagram  is relevant for both
neutrons and protons.
In the lower panel of fig. \ref{rb74.fig} the rotational
spectrum of  $^{74}_{37}$Rb$_{37}$ is constructed from the qp Routhians.
The lowest $T=0$ configurations are generated by exciting a 
quasi proton  and a quasi neutron.
The first is the positive parity odd spin band $[Aa]_0$.
 Next, $[Ab]_0$ and $[Ae]_0$  are expected. 
 As discussed in section \ref{t=0.sec},
the condition $T_y=0$ permits only one linear combination 
 of the two excitations, obtained by exchanging the quasi
proton with the quasi neutron, which we arbitrary label by
only one of the terms in order to keep the notation simple.
The lowest $T=1$ bands are generated by cranking in isospace,
using the realistic deformed Nilsson potential (for details see
\cite{pncsm}).
The lowest band is the vacuum $[0]_1$. It 
is crossed by the $T=0$ band $[Aa]_0$,
which has a large alignment. The crossing frequency 
is fairly well reproduced. Thus it seems, that this crossing is 
a phenomenon belonging to the realm of $t=1$ pair-correlations. 
  
\begin{figure}[t]
\mbox{\psfig{file=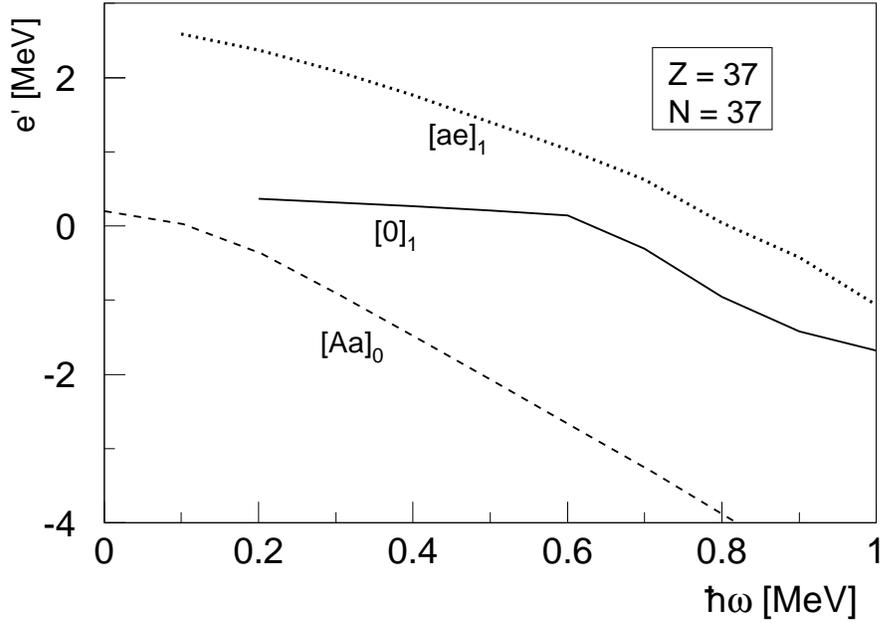,width=14cm}}
\caption{\label{rb74trs.fig}
Total routhians for  $^{74}_{37}$Rb$_{37}$ calculated by
means of the deformation optimized Woods Saxon Strutinsky method.  
 The text explains how the energy of the $T=1$ bands  relative
to the  energy of the $T=0$ ground state is fixed.
The parity and signature assignments $(\pi,\alpha)$ are:
Full lines (+,0), dashed (+,1) and
dotted (-,1). A Harris reference is subtracted.
   }
\end{figure}

The CSM assumptions of fixed deformation and pairing are too inaccurate
for the high frequency region. Of course one may combine
  the concept isorotation 
 with a more sophisticated mean-field calculations.
Fig. \ref{rb74trs.fig}  presents the spectrum of $^{74}$Rb as an example. 
Only pp- and nn- pairing is
considered, but in addition to the  monopole a quadrupole pair-field is
taken into account. For each configuration and frequency $\omega$,
 the deformation parameters are
individually optimized.  The calculations of  \cite{kr74} for the 
yrast sequence in $^{74}$Kr are used  for the configuration $[0]_1$ and
the results of an analogous TRS calculation \cite{rb74trs}
 for $^{74}$Rb are used for the configurations $[Aa]_0$ and $[Ae]_0$.
The relative energy of the $T=0$ and $T=1$ bands is calculated 
by setting at $\omega=0$ the energy difference between the
configurations $[0]$  in $N=38, ~Z=36$ and $N=37, ~Z=37$ equal to
the experimental value  for the isorotational energy.                
The same Harris reference as used for the experimental Routhians
 is subtracted from the calculated ones.  
 The calculated spectrum now agrees rather well with the data at 
high $\omega$.

\section{The $t=0$ pair-field}\label{t0}

Since the $t=0$ pair-field is an isoscalar
\begin{equation}  
[\vec T, \kappa_M]=0,
\end{equation}
and the qp Routhian  conserves the isospin. The qp
operators have $t=1/2$ and either $t_z= 1/2$ (neutron + proton hole) or
-1/2 (proton + neutron hole). The qp vacuum has $T=0$. The
one-qp excitations have $T=\pm 1/2$. The two-qp excitations can be combined
into $T=1,T_z=-1,0,1$ states,  analogous to eq. (\ref{t1field}), and
into $T=0$ states, which correspond to the odd linear combination
of the $T_z=0$ pairs.

The field conserves the parity of the total number of particles
\begin{equation}\label{coma}  
e^{-i(N+Z)\pi} \kappa_M e^{i(N+Z)\pi}=\kappa_M.
\end{equation}
This means that even-$A$ nuclei and odd-$A$ nuclei correspond to
different qp \confs (even or odd number of \qps) and have different
excitation spectra.
However, it does not conserve the neutron or proton number parity 
separately 
\begin{equation}\label{comn}  
e^{-iN\pi}\kappa_Me^{iN\pi}=e^{-iZ\pi}\kappa_Me^{iZ\pi}=-\kappa_M.
\end{equation}
This means that the same qp \conf with zero or
an even number of excited qps represents both even-even and odd-odd
nuclei. The wave function is a linear composition of 
states of  even-even and odd-odd particle numbers. 
Adjacent even-even and odd-odd $N=Z$ nuclei should have similar
excitation spectra.

In order to understand this  statement better,
let us briefly return to the familiar case of the pure neutron pair-field. 
For a \conf with an even number of qps the wave function is 
a linear combination of states with even particle number. It may be viewed
as the intrinsic state of a pair-rotational band. The different members
of the band correspond to $N, ~N\pm2, ~N\pm4, ..$. In the limit of very
strong breaking of the particle number $N$ conservation, all the 
members of the band have the same excitation spectrum, which is given by
the different intrinsic states. This situation is reached for  a very strong
pair-field, when $\Delta$ is much larger than the distance between 
the single-particle levels. In reality, $\Delta$ is only somewhat larger 
than the level distance (about 3 times in the heavy nuclei) and
the pair-correlations do not completely smear out the region near
the Fermi surface. Still one observes quite a remarkable similarity between
the spectra of the adjacent even- even nuclei.  For example, the first excited
two-qp state appears always at about $2\Delta$.

For the $t=0$ pair-field
the situation is analogous. However, adding  proton-neutron pairs,
brings us  from an even-even nucleus to an odd-odd one and the again to an 
even-even one, etc. This means, if $\Delta$ is larger than the  distance 
between the single-particle levels, the spectra of neighboring   
even-even nucleus and odd-odd $N=Z$ nuclei must be similar. This
represents a clear evidence, which can be checked experimentally.

The $t=0$  field consists of $J=1$  pairs, which
have the three  \am projections   $M=-1$, 0, and 1. 
Let us assume that $M$ is the projection on the x-axis,
which is the axis of rotation. The components have different signature
\begin{equation}\label{coms}  
{\cal R}_x(\pi) \kappa_M {\cal R}_x^{-1}(\pi)=(-)^M\kappa_M.
\end{equation}
Let us consider the important case that the deformed potential
$\Gamma$ (cf. eq. (\ref{hfbg})) conserves the signature.
If there is a $M=0$ field the qp Routhian conserves the signature
if there are $M=\pm 1$ fields the signature is not conserved. 

\subsection{The $M=0$ field}

Let us first consider the case $M=0$. It was first
discussed by Goswami and Kisslinger \cite{goswami}. It corresponds 
to the   ( $\alpha=m, \bar \alpha=-m, t=0$) pairs in Goodman's classification
(cf. his lecture and \cite{goodman}).  He finds for the even-even 
$N=Z$ nuclei with  $ 76\le A\ < 90 $ a $t=1$ pair-field at low spin.
 This is  in accordance with the our  discussion of the spectra in
 section \ref{t0}.
The observed spectra of adjacent even-even and odd-odd nuclei 
up to mass 80 are distinctly different. This excludes  a $t=0$ pair-field
with $\Delta$ larger than the single particle level distance. 
As discussed before, the different spectra of the even-even
and odd-odd  $N=Z$ nuclei can easily be understood in assuming
a $t=1$ pair-field.

 Goodman   finds a $t=0$ pair-field  of 
the ($\alpha, \bar \alpha$)-type for   $^{92}$Pd
\cite{goodman}. The potential $\Gamma$ is near spherical. The 
chemical potential is situated in the $g_{9/2}$ shell, which is 
almost degenerate. In such a situation one can expect that the spectra of
the even-even and odd-odd neighbors are similar. 
It would be interesting to see if the experiment confirms this.

Goswami and Kisslinger \cite{goswami} considered the possibility that
$\Delta$ is much smaller than the level distance $d$. 
In such a case, the gap between
the ground state and the first excited state is much
 smaller in the odd-odd nucleus 
(2$\Delta$) than in the even-even (2$\sqrt{\Delta^2+(d/2)^2}$).  
It is difficult to derive information about such a weak 
pair-field from the spectra. Similarly, it seems hard to
extract from the spectra clearcut evidence for a
mixed pair-field composed of $t=1$ and $t=0$ components, which
Goodman obtains for one set of input parameters \cite{goodman}.

\subsection{The $M=1$ field}

This type of pair-field is favored by rotation, because it carries
\am. One may speculate that such a field appears 
at high spin, where the rotation has destroyed the $t=1$ pairing and
made the phase space available for the new type of correlation. 
 
The $M=1$ pair-field has a special symmetry. It is odd under 
$e^{-iN\pi}$  (cf. (\ref{comn})) and under ${\cal R}_x(\pi)$
 (cf. (\ref{coms})). Therefore it is even under the 
combination of both and the qp Routhian is invariant,
\begin{equation}  
{\cal S}_N{\cal H}'{\cal S}_N^{-1}={\cal H}',
~~ {\cal S}_N=e^{-iN\pi}{\cal R}_x(\pi). 
\end{equation}
There is an analogy to reflection asymmetric 
axial nuclei \cite{simplex1,simplex2}.
Although both ${\cal R}_x(\pi)$ and the space inversion ${\cal P}$
do not leave the qp Routhian invariant, the combination 
${\cal S}= {\cal P}{\cal R}_x(\pi)$ does, which implies
the quantum number simplex. The bands are 
$\Delta I =1$ sequences of alternating parity.
The simplex determines  which parity belongs to $I$.
In the same way  ${\cal S}_N$ implies the quantum number $\gamma$
\begin{equation}\label{g}  
{\cal S}_N|\rangle=e^{-i\gamma\pi}|\rangle,
\end{equation}
which takes the values 0 and 1 for even $A$ and $\pm1/2$ for odd $A$.
Refering to the analogy with the simplex we suggest the name 
gauge-simplex or shorter gaugeplex. It relates the 
parity of the neutron number (or proton number) with the signature $\alpha$,   
\begin{equation}\label{gin}  
 e^{-i(I+N)\pi}=e^{-i\gamma\pi}.
\end{equation}
Here we used $ e^{-i\alpha\pi}= e^{-iI\pi}$.
A strong $M=1$ pair-field  implies that adjacent
even-even and odd-odd nuclei  join into a pair-rotational band 
of fixed gaugeplex. This means that the even-$I$ states become similar
to the odd-$I$ states of the neighbor and vice versa.

\begin{figure}[t]
\vspace*{-2cm}
\mbox{\psfig{file=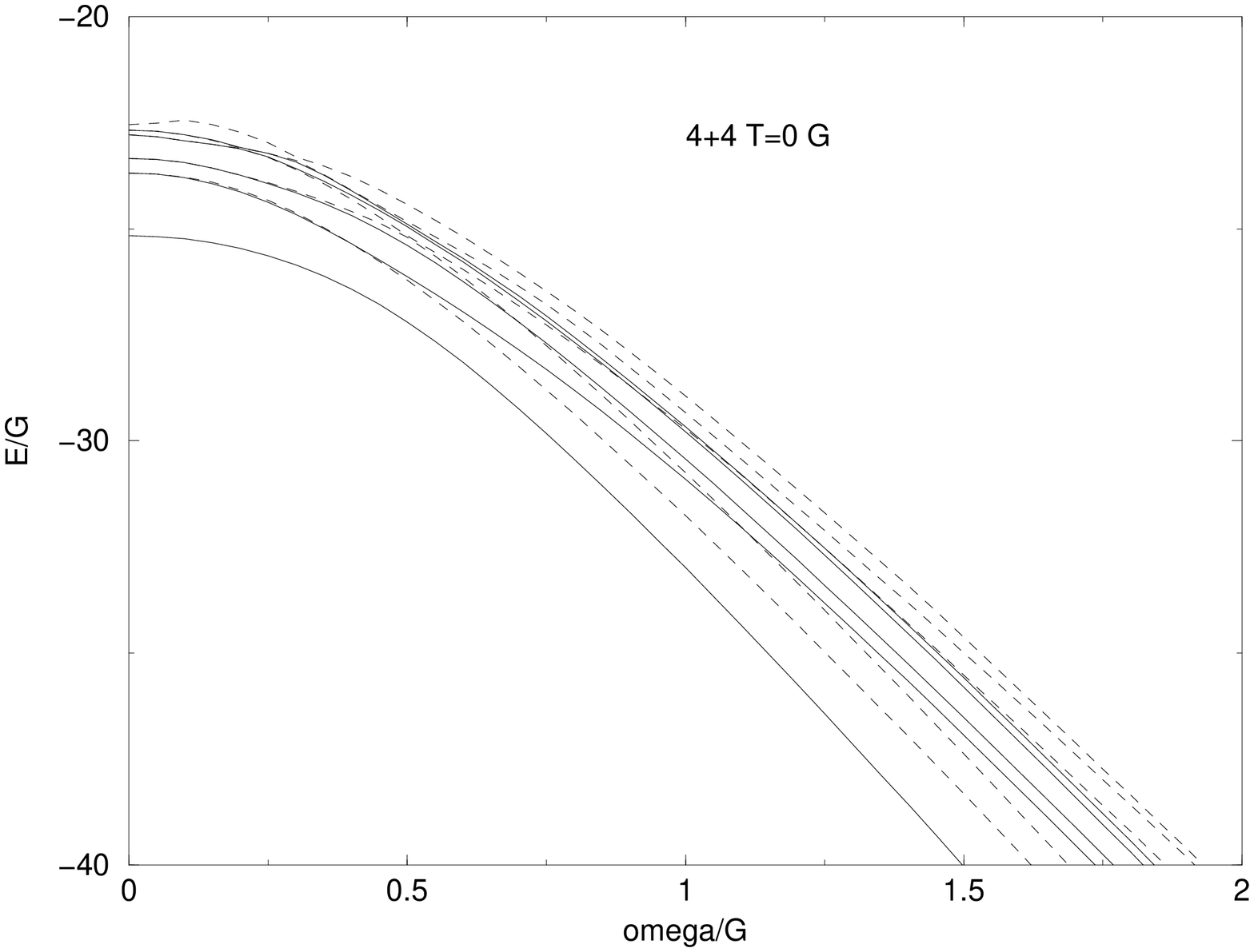,width=12cm,angle=-90}}
\vspace*{3cm}
\caption{\label{pn44.fig}
Total routhians of the $(Z=N=4)$ system as obtained by the shell model for a 
$t=0$ $\delta$ -interaction.
Full lines correspond to even spins and dashed ones to odd spins. 
   }
\end{figure}

\begin{figure}[t]
\vspace*{-2cm}
\mbox{\psfig{file=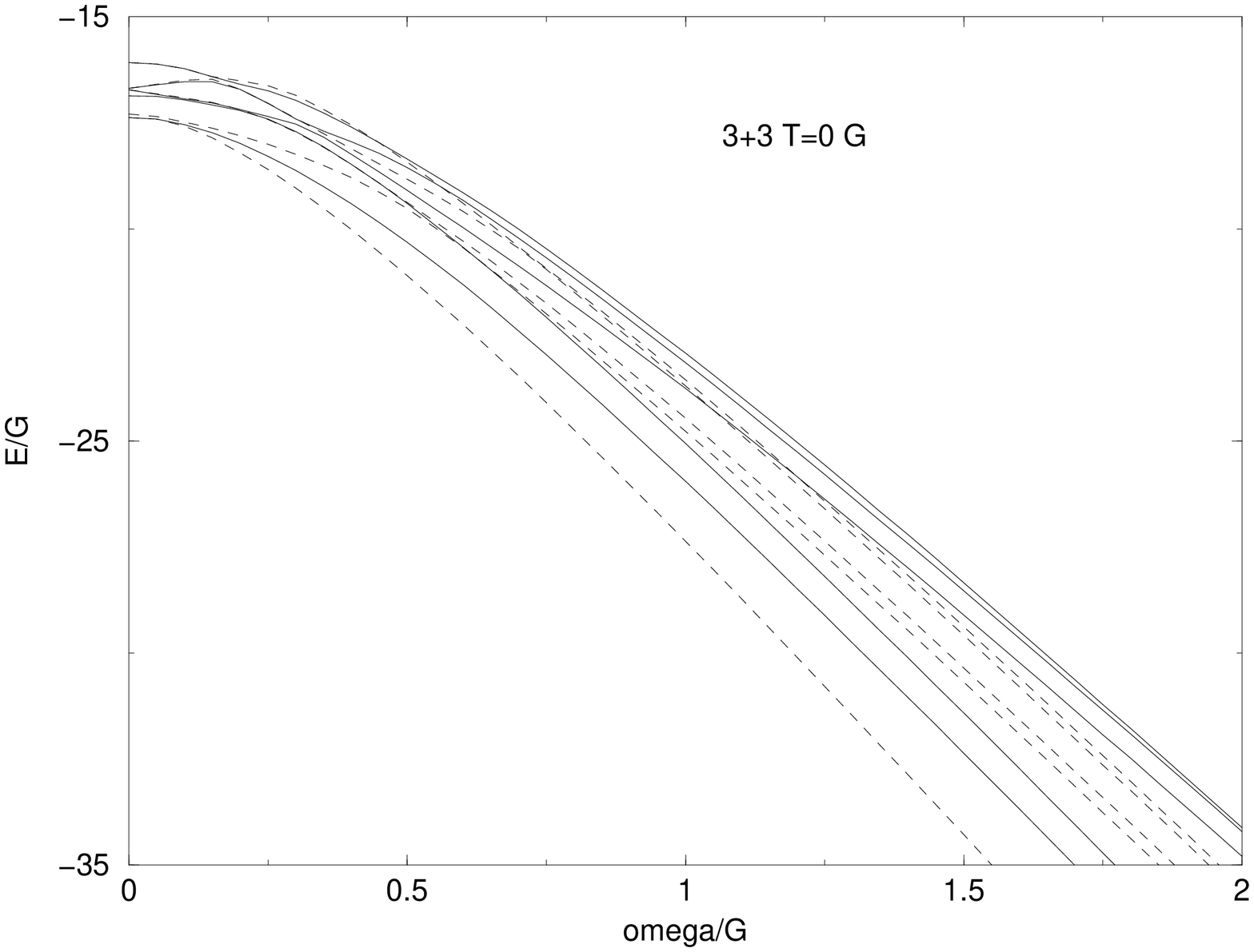,width=12cm,angle=-90}}
\vspace*{3cm}
\caption{\label{pn33.fig}
Total Routhians of the $(Z=N=3)$ system obtained by the shell model for a 
$t=0$ $\delta$ -interaction.
Full lines correspond to even spins and dashed ones to odd spins. 
   }
\end{figure}

In order to investigate this interesting structure we carried out 
shell model calculations for our study case of a $f_{7/2}$-shell. In order to
enhance the $t=0$ correlations we modified the interaction. Only the 
odd-$J$ multipoles of the $\delta$-interaction, which
have $t=0$,  were taken into account  ($t=0$ $\delta$ -interaction). 
 Figs. \ref{pn44.fig} and
\ref{pn33.fig} show the results. It is seen that for $\omega/G > 1$ the
spectra of the systems $Z=N=4$ and $Z=N=3$ become very similar if the 
states of opposite signature are compared. 
The similarity may be ascribed to a gradual built up of the $t=0, ~M=1$
correlations caused by the increasing frequency. This interpretation
is supported by the development of a gap between the 
yrast and yrare state. At $\omega/G=2$, the  distances between  the lowest
Routhians are in units of $G$: 2.4, 1.0, 1.2 for $Z=N=3$ and 2.2, 0.9, 0.9
for $Z=N=4$. So far we have not been able to find a self-consistent  
pair-field in this frequency region. It is not clear at this point 
what this means. The work is very recent. It may be that we just did
not use the right initial wavefunction for the iterative solution of the
HFB equations. It is possible that the pair-correlations are 
of vibrational type and there is no static HFB solution.
But it could also be that there is a completely different
explanation for the similarity.

\section{Conclusions}

The spectra of $N \approx Z$ nuclei in the region $40<A<80$ are 
consistent with a $t=1$ pair-field at low
and moderate spin (cf. S. Lenzi's
lecture at this meeting). This conclusion is in accordance with 
the analysis of two-particle transfer data,
which has already provided strong 
independent  evidence for this type of pair-field \cite{pairvib}.  

The $t=1$ pair-field breaks the isotropy with respect to rotations in 
isospace. Therefore, the mean field solutions must be interpreted
as intrinsic states of pair-rotational bands. This has two consequences.

As intrinsic state, one may use the orientation in isospace with a zero 
proton-neutron pair-field. The intrinsic spectrum looks as if there was no
such field. Only for $N=Z$ certain excitations ($T_y\not=0$) are forbidden.
The proton-neutron pair-correlations appear via the restoration of the
isospin in the total wavefunction.   
  
On top of each intrinsic state, there is a pair-rotational band.
However, these additional states with $T>T_z$ lie high in energy. 
Only in the odd-odd $N=Z$ nuclei the first excited pair-rotational
state based on the intrinsic ground state has a similar energy as the first
excited intrinsic state. Hence the
 appearance of this low lying $T=1$ rotational
band is a consequence of the $t=1$ pair-correlations.

The  $t=0$ pair-field, which  has  other symmetries than the
$t=1$ field, leads to a different pattern of  excited states.
Since the pairs carry finite \am ($J=1$ is expected to be most important)
one must distinguish between the fields with signature 0 and 1.
 
A substantial  pair-field with signature 0 would  
show up as similar spectra in adjacent even-even and odd-odd $N=Z$ nuclei.
Such a  similarity is not seen in the experimental spectra. 
It should appear around $A=92$, where Goodman \cite{goodman}
predicts this type of pair-field at low spin.

The pair-field with signature 1 is favored by rotation. It may appear
at high spin. A substantial field of this type would show up as
 similar spectra in adjacent  $N=Z$ nuclei,
provided the odd/even spins in  even-even nucleus 
are compared with the even/odd spins in the  odd-odd nucleus.     

\section{Acknowledgment}
The work was supported by the grant DE-FG02-95ER40934.

\newpage
\onecolumn

\newpage

\newpage

\newpage

\newpage

\newpage

\newpage

\newpage
\newpage

\end{document}